\documentclass[article,aps,showpacs,amsfonts,epsf]{revtex4}

\usepackage{graphicx}

\newcommand{\E}{{\mathcal{E}}}
\newcommand{\s}{\sigma}

\renewcommand{\a}{\alpha}

\newcommand{\be}{\begin{equation}}
\newcommand{\ee}{\end{equation}}
\newcommand{\bea}{\begin{eqnarray}}
\newcommand{\eea}{\end{eqnarray}}
\newcommand{\ba}{\begin{array}}
\newcommand{\ea}{\end{array}}

\def\J#1#2#3#4{{#1} {#2} (#3) #4}
\def\PRD{Phys. Rev. D}

\def\PRL{Phys. Rev. Lett.}
\def\PTP{Prog. Theor. Phys.}

\def\APL{Ann. Phys. (Leipzig)}

\def\JMP{J. Math. Phys.}

\def\CQG{Class. Quantum Grav.}

\def\PLB{Phys. Lett. B}
\def\PTP{Prog. Theor. Phys.}

\def\NPB{Nucl. Phys. B}

\def\ib{{\it ibid.}}

\begin{document}
\draft
\title{A note on static dyonic diholes}

\author{H. Garc\'ia-Compe\'an, V. S. Manko}
\address{Departamento de F\'\i sica, Centro de Investigaci\'on y de
Estudios Avanzados del IPN,\\ A.P. 14-740, 07000 M\'exico D.F.,
Mexico}

\begin{abstract}
In this brief note we argue that a dyonic generalization of the
Emparan-Teo dihole solution is described by a static diagonal
metric and therefore, contrary to the claim made in a recent paper
by Cabrera-Munguia et al., does not involve any ``non-vanishing
global angular momentum'' and rotating charges.
\end{abstract}


\maketitle


\section{Introduction}

In a recent paper \cite{CLM}, Cabrera-Munguia et al. presented an
exact stationary axisymmetric solution of the Einstein-Maxwell
equations for two unequal dyons and considered some of its
limiting cases. In subsection 5.3 they assert that in the limit of
vanishing total angular momentum, their binary dyonic
configuration still has ``non-vanishing global angular momentum''
and, moreover, ``the magnetic monopole charges arise from the
rotation of the Reissner-Nordstr\"om black holes''. Our note aims
at demonstrating explicitly that the dyonic generalizations of the
symmetric \cite{ETe} and asymmetric \cite{MRS} dihole spacetimes
(to which supposedly must lead that limit) are static
intrinsically, so that the claim by Cabrera-Munguia et al. is
erroneous and misleading.

\section{The dyonic Emparan-Teo dihole}

The static dihole solution describing two Reissner-Nordstr\"om
black holes \cite{Rei,Nor} with equal masses and opposite electric
charges was obtained by Emparan and Teo \cite{ETe} as a particular
specialization of the double-Kerr-Newman solution \cite{MMR}. The
physical parametrization of the Emparan-Teo dihole and its
magnetostatic analog were later worked out in \cite{CGM}, the two
static versions of the dihole spacetime being related by Bonnor's
theorem \cite{Bon}. The dyonic generalization of the Emparan-Teo
dihole is obtainable from the original electrostatic solution by
means of the duality rotation, and below we write down the
corresponding expressions of the Ernst potentials \cite{Ern},
satisfying the equations (for real $\E$)
\bea (\E + \bar\Phi\Phi) \nabla^{2}\E & = &
(\nabla\E + 2 \bar\Phi\nabla\Phi) \cdot \nabla\E, \nonumber\\
(\E + \bar\Phi\Phi) \nabla^{2}\Phi & = & (\nabla\E + 2
\bar\Phi\nabla\Phi) \cdot \nabla\Phi, \label{Eeq} \eea
which arise as a by-product of the recent paper \cite{MRa} on
stationary diholes:
\bea \E&=&\frac{A-B}{A+B}, \quad \Phi=\frac{{\cal Q}C}{A+B}, \nonumber\\
A&=&R^2(M^2-|{\mathcal Q}|^2\tau)(R_+-R_-)(r_+-r_-)+
4\s^2(M^2+|{\mathcal Q}|^2\tau)(R_+-r_+)(R_--r_-)
\nonumber\\ &&+2R^2\s^2(R_+r_-+R_-r_+), \nonumber\\
B&=&2MR\s[(R\s-2M^2)(R_++r_-)+(R\s+2M^2)(R_-+r_+)],
\nonumber\\
C&=&\frac{2R\s(R-2M)}{R^2-4\s^2}[(R-2\s)(R\s+2M^2)(R_+-r_-)
+(R+2\s)(R\s-2M^2)(R_--r_+)], \nonumber\\
R_\pm&=&\sqrt{\rho^2+\left(z+\frac{1}{2}R\pm\sigma\right)^2},
\quad r_\pm=\sqrt{\rho^2+\left(z-\frac{1}{2}R\pm\sigma\right)^2}.
\label{EF} \eea
In the above formulae
\bea {\cal Q}&=&Q+i{\cal B}, \quad |{\mathcal Q}|^2=Q^2+{\cal
B}^2,
\nonumber\\
\s&=&\sqrt{M^2-|{\mathcal Q}|^2\frac{R-2M}{R+2M}}, \quad
\tau=\frac{R^2-4M^2}{(R+2M)^2+4|{\mathcal Q}|^2}, \label{sig} \eea
the real parameters $M$, $Q$, ${\cal B}$ and $R$ being,
respectively, the mass, electric charge, magnetic charge of the
upper dyon and the separation distance; the characteristics of the
lower constituent are $M$, $-Q$, $-{\cal B}$. The Weyl cylindrical
coordinates $(\rho,z)$ enter the expressions (\ref{EF}) only
through the functions $R_\pm$ and $r_\pm$.

By setting ${\cal B}=0$ in (\ref{EF}) and (\ref{sig}), one comes
to the Ernst potentials of the Emparan-Teo electrostatic solution
\cite{ETe}, and the limit $Q=0$ leads to the magnetostatic analog
of the Emparan-Teo dihole \cite{CGM}. Note, that the function $\E$
in (\ref{EF}) is real, while the electromagnetic potential $\Phi$
represents a product of the complex constant ${\cal Q}$ and a real
function; therefore, the metric defined by these $\E$ and $\Phi$
remains static and diagonal. Below we give the corresponding
metric functions $f$ and $\gamma$ from the Weyl line element
\be d s^2=f^{-1}[e^{2\gamma}(d\rho^2+d z^2)+\rho^2d\varphi^2]-fd
t^2, \label{Weyl} \ee
together with the electric $A_t$ and magnetic $A_\varphi$
components of the electromagnetic 4-potential,
\bea f&=&\frac{A^2-B^2+|{\mathcal Q}|^2 C^2} {(A+B)^2}, \quad
e^{2\gamma}=\frac{A^2-B^2+|{\mathcal Q}|^2 C^2}
{16R^4\s^4R_+R_-r_+r_-}, \quad A_t=-\frac{QC}{A+B}, \quad
A_\varphi=\frac{{\mathcal B}(I-zC)}{A+B},
\nonumber\\
I&=&-\frac{2M(R-2M)}{R^2-4\s^2}[2R^2(M^2-\s^2)(R_+r_++R_-r_-)
+2\s^2(R^2-4M^2)(R_+R_-+r_+r_-)] \nonumber\\ &&+(R-2M)
\{2M[R\s(R_+r_--R_-r_+) +2M^2(R_+r_-+R_-r_+)] \nonumber\\
&&+R\s[R\s(R_++R_-+r_++r_-) +6M^2(R_+-R_--r_++r_-) +8MR\s]\},
\label{mf} \eea
and one can see in particular that $A_\varphi$ is defined by a
more concise expression than in the paper \cite{CGM}. Here it is
worth noting that the duality rotation of the potential $\Phi$
leaves the metric unchanged, and so the transformation
\be \Phi'=e^{i\alpha_0}\Phi, \quad \tan\a_0=-{\cal B}/Q,
\label{dr} \ee
would convert the solution (\ref{EF}) into the electrostatic one
(${\rm Im}\,\Phi'=0$) described by the Weyl metric with the
functions $f$ and $\gamma$ from (\ref{mf}), thus confirming the
staticity of the dyonic Emparan-Teo solution (\ref{EF}).

The absence of any stationary energy flows created by the electric
and magnetic charges in the binary dyonic configuration (\ref{EF})
can be easily established by analyzing the associated Poynting
vector. In \cite{HGP} it was demonstrated  that the Poynting
vector of a stationary axisymmetric electrovac spacetime can have
only one non-zero component, and in \cite{MRSS} this
$\varphi$-component was shown to be defined by the following
simple formula:
\be S^\varphi=\frac{\sqrt{f}{\rm e}^{-2\gamma}}{4\pi\rho}{\rm
Im}(\bar\Phi_{,\rho}\Phi_{,z}). \label{Poy} \ee
Then, taking into account that $\bar\Phi_{,\rho}\Phi_{,z}$ is a
real function in the case of the solution (\ref{EF}), one
immediately gets $S^\varphi=0$, which means the absence of
frame-dragging effects due to electromagnetic field in the metric
(\ref{mf}). The latter metric is therefore static intrinsically.

Let us also mention for completeness  that on the upper and lower
horizons ($\rho=0$, ${\textstyle\frac12}R-\s\le z\le
{\textstyle\frac12}R+\s$ and $\rho=0$,
$-{\textstyle\frac12}R-\s\le z\le -{\textstyle\frac12}R+\s$,
respectively) the surface gravity $\kappa^H$ and horizon's area
$S^H$ of the dyonic Emparan-Teo solution are given by the
expressions
\be \kappa^H=\frac{R\s(R+2\s)}{(R+2M)^2(M+\s)^2}, \quad
S^H=\frac{4\pi(R+2M)^2(M+\s)^2}{R(R+2\s)}, \label{kap} \ee
while the Ernst potential $\Phi$ on the upper horizon takes the
complex constant value
\be \Phi^H_{ext}=\frac{{\cal Q}(M-\s)}{|{\cal Q}|^2} \label{FH}
\ee
(on the lower horizon, $\Phi^H_{ext}$ changes its sign), thus
representing a complex extension of the electric potential
$\Phi^H$ from the Smarr mass formula \cite{Sma}. The generalized
mass relation involving the electric and magnetic charges has the
form
\be M=\frac{1}{4\pi}\kappa^H S^H+\bar{\cal Q}\Phi^H_{ext},
\label{Sma_mod} \ee
and it is verified identically on both horizons since the complex
charge ${\cal Q}$, similar to $\Phi^H_{ext}$, changes its sign on
the lower horizon.

\section{The asymmetric dyonic dihole}

The case of the dyonic asymmetric dihole solution is fully
analogous to the previous case of the dyonic Emparan-Teo dihole.
The Ernst potentials of that solution are obtainable by means of
the substitutions $Q\to{\cal Q}$, $Q^2\to|{\cal Q}|^2$ from the
formulae of the paper \cite{MRS} defining the respective
potentials of the asymmetric electric dihole, thus eventually
leading to the metric functions $f$ and $\gamma$ in the Weyl line
element (\ref{Weyl}) and to the electric $A_t$ and magnetic
$A'_\varphi$ potentials of the form
\bea f&=&\frac{A^2-4B^2+4|{\cal Q}|^2C^2}{(A+2B)^2}, \quad
e^{2\gamma}=\frac{A^2-4B^2+4|{\cal Q}|^2C^2}{K_0^2 R_+R_-r_+r_-},
\quad A_t=-\frac{2QC}{A+2B}, \quad A'_\varphi=\frac{2{\cal
B}C}{A+2B},
 \nonumber\\
A&=&\Sigma\,\s[\nu(R_++R_-)(r_++r_-)+4\kappa(R_+R_-+r_+r_-)]
-(|{\cal Q}|^2\mu^2\nu-2\kappa^2)(R_+-R_-)(r_+-r_-), \nonumber\\
B&=&\Sigma\,\s[(m\nu +2M\kappa)(R_++R_-)+(M\nu+2m\kappa)(r_++r_-)]
+|{\cal Q}|^2(\mu-\mu^2)(\nu-2\kappa)
\nonumber\\ &&\times[\s(R_+-R_-)-\Sigma(r_+-r_-)]
-2R\kappa[M\s(R_+-R_-)-m\Sigma(r_+-r_-)], \nonumber\\
C&=&\Sigma\,\s(1-\mu)(\nu-2\kappa)(r_++r_--R_+-R_-)
+\s[M\mu\nu +2\kappa(m\mu-R+R\mu)](R_+-R_-) \nonumber\\
&&+\Sigma[m\mu\nu+2\kappa(M\mu-R+R\mu)](r_+-r_-), \nonumber\\
K_0&=&4\Sigma\,\s[R^2-(M-m)^2+4|{\cal Q}|^2(1-\mu)^2],
\nonumber\\
R_\pm&=&\sqrt{\rho^2+\left(z+\frac{1}{2}R\pm\Sigma\right)^2},
\quad r_\pm=\sqrt{\rho^2+\left(z-\frac{1}{2}R\pm\s\right)^2},
\label{fg_as} \eea
where
\bea {\cal Q}&=&Q+i{\cal B}, \quad |{\mathcal Q}|^2=Q^2+{\cal
B}^2, \nonumber\\
\Sigma&=&\sqrt{M^2-|{\cal Q}|^2(1-2\mu)}, \quad
\s=\sqrt{m^2-|{\cal Q}|^2(1-2\mu)}, \nonumber\\
\mu&=&\frac{M+m}{R+M+m}, \quad \kappa=Mm+|{\cal Q}|^2(1-\mu)^2,
\quad \nu=R^2-M^2-m^2+2|{\cal Q}|^2(1-\mu)^2, \label{sig_as} \eea
and now the lower dyonic black hole with the horizon $\rho=0$,
$-{\textstyle\frac12}R-\Sigma\le z\le
-{\textstyle\frac12}R+\Sigma$ has the mass $M$, electric charge
$Q$ and magnetic charge ${\cal B}$, whereas the mass and charges
of the upper dyonic constituent with the horizon $\rho=0$,
${\textstyle\frac12}R-\s\le z\le {\textstyle\frac12}R+\s$ are $m$,
$-Q$ and $-{\cal B}$ (see Fig.~1). Note that in formulae
(\ref{fg_as}) we give the expression of the magnetic potential
$A'_\varphi$ (the imaginary part of the Ernst potential $\Phi$)
instead of $A_\varphi$ because the latter has a rather cumbersome
form that can be inferred from Eq.~(28) of \cite{MRS}.

\begin{figure}
  \centering
    \includegraphics[width=12cm]{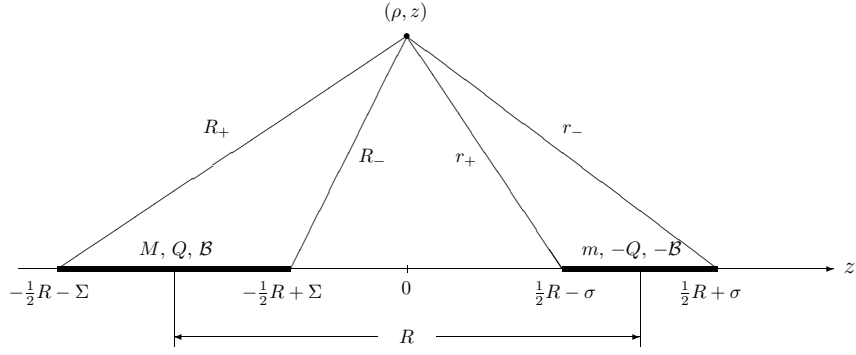}
  \caption{Location of
the dyonic black-hole constituents on the symmetry axis.}
  \label{fig1}
\end{figure}

There can be no doubt that the spacetime of the asymmetric dyonic
dihole (\ref{fg_as}) is static intrinsically: it is described by a
diagonal static metric and, as can be easily checked with the aid
of the formula $\Phi=-A_t+iA'_\varphi$, the $\varphi$-component of
the Poynting vector (\ref{Poy}) is zero, thus proving the absence
of any frame-dragging phenomena in this spacetime.

Let us also mention that the non-equal dyonic black-hole
constituents verify the generalized Smarr formula (\ref{Sma_mod}).
Thus, for instance, restricting ourselves to the lower
constituent, we will have
\be \kappa^H=\frac{\Sigma[(R+\Sigma)^2-\s^2]}
{(R+M+m)^2(M+\Sigma)^2}, \quad
S^H=\frac{4\pi(R+M+m)^2(M+\Sigma)^2} {(R+\Sigma)^2-\s^2}, \quad
\Phi^H_{ext}=\frac{{\cal Q}(1-2\mu)}{M+\Sigma}, \label{kap_as} \ee
with which the relation (\ref{Sma_mod}) holds identically.

\section{Concluding remarks}

Therefore, our analysis makes it very clear that the dyonic
generalizations of the known electrostatic solutions for black
diholes \cite{ETe,MRS} are static as well, despite the presence in
them of both electric and magnetic charges. The erroneous physical
interpretations of the dyonic configurations made in the paper
\cite{CLM} spring from improper use by Cabrera-Munguia et al. of
Tomimatsu's formula for the angular momentum \cite{Tom} that
forced those authors to make some incorrect redefinitions (see
\cite{MSa,MGa} for details). We hope that our results may be
helpful in the future for constructing more general static
spacetimes, for instance in the presence of a dilaton field.

\section*{Acknowledgments}

This work was partially supported by Project~128761 from CONACyT
of Mexico.

\newpage


\begin{references}
\bibitem{CLM} I. Cabrera-Munguia, C. L\"ammerzahl, A. Mac\'ias,
\J{\PLB}{743}{2015}{357}.

\bibitem{ETe} R. Emparan, E. Teo, \J{\NPB}{610}{2001}{190}.

\bibitem{MRS} V.S. Manko, E. Ruiz, J. S\'anchez-Mondrag\'on, \J{\PRD}{79}{2009}{084024}.

\bibitem{Rei} H.~Reissner, \J{\APL}{355}{1916}{106}.

\bibitem{Nor} G.~Nordstr\"om, Proc. K. Ned. Akad. Wet. 20 (1918) 1238.

\bibitem{MMR} V.S. Manko, J. Mart\'in, E. Ruiz, \J{\JMP}{35}{1994}{6644}.

\bibitem{CGM} J.A. C\'azares, H. Garc\'ia-Compe\'an, V.S. Manko,
\J{\PLB}{662}{2008}{213}; \J{\ib}{665}{2008}{426} (Erratum).

\bibitem{Bon} W.B. Bonnor, Proc. Phys. Soc. Lond. A 67 (1954) 225.

\bibitem{Ern} F.~J.~Ernst, Phys. Rev. 168 (1968) 1415.

\bibitem{MRa} V.S. Manko, R.I. Rabad\'an, J.D. Sanabria-G\'omez, \J{\PRD}{89}{2014}{064049}.

\bibitem{HGP} L. Herrera, G.A. Gonz\'alez, L.A. Pach\'on, J.A.
Rueda, \J{\CQG}{23}{2006}{2395}.

\bibitem{MRSS} V.S.~Manko, E.D.~Rodchenko, B.I.~Sadovnikov,
J.~Sod-Hoffs, \J{\CQG}{23}{2006}{5389}.

\bibitem{Sma} L. Smarr, \J{\PRL}{30}{1973}{71}.

\bibitem{Tom} A. Tomimatsu, \J{\PTP}{72}{1984}{73}.

\bibitem{MSa} V.S. Manko, J.D. Sanabria-G\'omez, \J{\PRD}{91}{2015}{088501}.

\bibitem{MGa} V.S. Manko, H. Garc\'ia-Compe\'an,
arXiv:1506.03870 [gr-qc].

\end{references}
\end{document}